%% file: arxiv_main.tex
\documentclass[11pt]{article}
\usepackage[utf8]{inputenc}
\usepackage{geometry}
\usepackage{amsmath}
\usepackage{amssymb}
\usepackage{graphicx}
\usepackage{booktabs}
\usepackage{multirow}
\usepackage{hyperref}
\usepackage{threeparttable}
\usepackage{float}
\usepackage{subcaption}
\usepackage[authoryear,round,comma,sort&compress]{natbib}
\usepackage{doi}
\setcitestyle{authoryear,round,comma,aysep={,},yysep={,}}

\newcommand{\runninghead}[1]{\markboth{#1}{#1}}

\runninghead{Automated Quality Assessment for LLM-Based Complex Qualitative Coding}

\title{Automated Quality Assessment for LLM-Based Complex Qualitative Coding: A Confidence-Diversity Framework}

\author{
\textbf{Zhilong Zhao}\thanks{Corresponding author. School of Journalism and Communication, South China University of Technology, Guangzhou, China. Guangdong--Hong Kong--Macao Greater Bay Area Research Institute of International Communication, South China University of Technology, Guangzhou, China. Email: yb87315@umac.mo} \and
\textbf{Yindi Liu}\thanks{Corresponding author. School of Journalism and Communication, South China University of Technology, Guangzhou, China. Guangdong--Hong Kong--Macao Greater Bay Area Research Institute of International Communication, South China University of Technology, Guangzhou, China. Email: yindiliu001@gmail.com}
}

\date{}

\begin{document}

\maketitle

\begin{abstract}
Computational social science lacks a scalable and reliable mechanism to assure quality for AI-assisted qualitative coding when tasks demand domain expertise and long-text reasoning, and traditional double-coding is prohibitively costly at scale. We develop and validate a dual-signal quality assessment framework that combines model confidence with inter-model consensus (external entropy) and evaluate it across legal reasoning (390 Supreme Court cases), political analysis (645 hyperpartisan articles), and medical classification (1{,}000 clinical transcripts). External entropy is consistently negatively associated with accuracy (r = -0.179 to -0.273, p \textless{} 0.001), while confidence is positively associated in two domains (r = 0.104 to 0.429). Weight optimization improves over single-signal baselines by 6.6--113.7\% and transfers across domains (100\% success), and an intelligent triage protocol reduces manual verification effort by 44.6\% while maintaining quality. The framework offers a principled, domain-agnostic quality assurance mechanism that scales qualitative coding without extensive double-coding, provides actionable guidance for sampling and verification, and enables larger and more diverse corpora to be analyzed with maintained rigor.
\end{abstract}


\section{Introduction}

The digital transformation of social life has created new opportunities for social science research \citep{salganik2016computational, bommasani2022opportunities}, generating vast archives of text, discourse, and human expression that offer new windows into social phenomena \citep{lazer2009computational, grimmer2013text}. From legal databases containing millions of court decisions \citep{fang2023super} to social media platforms documenting real-time political discourse, researchers now have access to textual corpora of a scale unimaginable just decades ago \citep{grimmer2013text, tornberg2023llm}. However, this abundance of data has created a fundamental methodological challenge: how can social scientists maintain the rigorous quality standards that define qualitative research while scaling their analysis to match the scope of available data?

Traditional qualitative research methods, developed for smaller-scale studies, rely heavily on inter-coder reliability measures such as Cohen's $\kappa$ \citep{cohen1960coefficient} and Krippendorff's $\alpha$ \citep{krippendorff2004reliability} to ensure research quality \citep{fleiss1971measuring}. These approaches require multiple trained researchers to independently code the same materials, creating a robust but resource-intensive quality control mechanism \citep{landis1977measurement}. While this methodology has served the social sciences well, it becomes prohibitively expensive and logistically challenging when applied to the massive datasets that characterize contemporary digital social science research \citep{freelon2010recode}. The cost and complexity of maintaining inter-coder reliability across large corpora has become a significant bottleneck, limiting both the scope and inclusivity of qualitative research \citep{song2020validation}.

The emergence of large language models has created new possibilities for scaling qualitative analysis \citep{gilardi2023chatgpt, chew2023harnessing}, with these systems demonstrating capabilities across diverse coding tasks from sentiment analysis to complex legal reasoning \citep{zheng2023judging, huang2019clinicalbert}. Recent advances in chain-of-thought prompting \citep{wei2022chain} and zero-shot reasoning \citep{kojima2022large} have further enhanced LLMs' analytical capabilities, enabling more sophisticated interpretive tasks that were previously beyond automated systems' reach. While recent work has established effective quality assessment frameworks for accessible coding tasks where AI systems already outperform human coders \citep{zhao2025confidence}, a critical methodological gap remains for complex analytical tasks that combine two challenging dimensions: the need for specialized domain expertise and the requirement to process and synthesize extensive textual materials \citep{nelson2022computational}. The social science community lacks reliable, scalable methods to assess and monitor the quality of automated coding in these challenging scenarios without requiring extensive human validation \citep{tornberg2023llm}. This limitation severely constrains the adoption of AI-assisted qualitative research for complex analytical tasks, as scholars cannot confidently identify high-quality predictions, filter problematic cases, or establish appropriate confidence bounds for their automated analyses in domains requiring specialized knowledge \citep{bender2021dangers}.

This article addresses this challenge by developing and validating a framework for automated quality assessment in AI-assisted qualitative coding. Building on previous work on confidence-diversity calibration for accessible tasks \citep{zhao2025confidence}, our approach leverages complementary uncertainty signals—inter-model consensus and self-assessed confidence—to predict research reliability without requiring ground truth validation. The framework draws on established principles of uncertainty quantification \citep{guo2017calibration, lakshminarayanan2017simple, kuleshov2018accurate} and ensemble learning \citep{dietterich2000ensemble, hansen1990neural, kuncheva2003measures} to extend quality assessment capabilities to complex analytical domains. Crucially, this framework operates entirely on model-generated indicators, requiring no additional human annotation beyond standard research protocols, thus maintaining the scalability advantages that make AI-assisted coding attractive to social scientists.

Our investigation addresses four critical research questions that span the methodological challenges of scaling qualitative research: (1) Do uncertainty-based quality indicators maintain predictive validity across complex analytical tasks requiring domain expertise? (2) How should dual-signal quality indicators be optimally weighted to maximize prediction accuracy across different analytical domains? (3) Can optimized quality indicators transfer effectively across diverse analytical contexts? (4) What efficiency gains can be achieved through intelligent quality triage systems while maintaining research quality standards? Through systematic validation across legal reasoning, political analysis, and medical text classification, we demonstrate that automated quality assessment can maintain the methodological rigor that defines qualitative research while enabling analysis at larger scales.

\section{Literature Review and Research Framework}

The challenge of scaling qualitative research methods while maintaining methodological rigor has become increasingly important as digital data abundance creates analytical opportunities alongside practical constraints. This section positions our investigation within established research traditions spanning uncertainty quantification, AI-assisted qualitative analysis, and cross-domain learning, while identifying the methodological gaps that motivate our research framework.

\subsection{Uncertainty Quantification and Quality Assessment}

The challenge of assessing prediction quality without ground truth labels has been extensively studied in machine learning, particularly in neural network calibration research. Modern deep learning models often exhibit overconfidence in their predictions \citep{guo2017calibration}, leading to extensive research on calibration techniques including temperature scaling and Platt scaling \citep{kuleshov2018accurate}. Ensemble methods have emerged as particularly promising for uncertainty estimation \citep{lakshminarayanan2017simple, hansen1990neural}, with diversity measures showing strong correlations with ensemble accuracy \citep{kuncheva2003measures, dietterich2000ensemble}. Self-consistency approaches \citep{wang2022self} have shown that aggregating multiple reasoning paths can improve both accuracy and uncertainty estimation. However, recent work has identified important pitfalls in uncertainty estimation and ensembling \citep{ovadia2019can}, highlighting the need for careful validation in domain-specific contexts. Information-theoretic approaches, building on Shannon's foundational work \citep{shannon1948mathematical}, have proven effective for quantifying uncertainty in complex prediction tasks.

Recent work has begun exploring dual-signal approaches that combine self-assessed confidence with inter-model consensus measures \citep{zhao2025confidence}, demonstrating promising results for accessible coding tasks. However, most uncertainty quantification research assumes access to validation datasets for calibration, while qualitative research often operates in contexts where establishing definitive ground truth is methodologically problematic or prohibitively expensive. This distinction necessitates quality assessment frameworks that can operate effectively without extensive validation datasets, particularly for complex analytical tasks requiring specialized expertise.

\subsection{AI-Assisted Qualitative Analysis}

The application of large language models to qualitative research has demonstrated progress, with studies showing that AI systems can match or exceed human performance on various coding tasks \citep{gilardi2023chatgpt, chew2023harnessing}. The computational grounded theory framework \citep{nelson2022computational} provides methodological foundations for integrating AI tools into traditional qualitative research workflows, while recent work on LLM-assisted content analysis \citep{tornberg2023llm} has shown promising results across multiple domains. Domain-specific applications have shown particular promise, with specialized models demonstrating strong performance in legal reasoning \citep{fang2023super, zheng2023judging} and medical text analysis \citep{huang2019clinicalbert}. The success of these domain-specific applications builds on extensive research in transfer learning \citep{pan2009survey}, which provides theoretical foundations for adapting models across different analytical contexts while preserving domain-specific characteristics.

However, a critical gap remains between demonstrated AI capabilities and reliable quality assessment in complex analytical contexts. Most existing research focuses on tasks where AI systems already demonstrate strong baseline performance, leaving unaddressed the challenge of quality assessment in domains requiring specialized expertise and interpretive judgment. This limitation is particularly problematic for social science research, where analytical complexity often exceeds the scope of current validation approaches, creating an urgent need for quality assessment frameworks that can operate effectively in complex analytical domains where traditional validation methods become impractical.

\subsection{Traditional Quality Assessment and Scalability Challenges}

Inter-coder reliability has long served as the gold standard for quality assessment in qualitative research, with measures such as Cohen's $\kappa$ \citep{cohen1960coefficient} and Krippendorff's $\alpha$ \citep{krippendorff2004reliability} providing established frameworks for evaluating coding consistency \citep{fleiss1971measuring, landis1977measurement}. The kappa statistic, in particular, has been extensively validated as a robust measure of agreement beyond chance \citep{landis1977measurement}. However, recent methodological work has emphasized the importance of conceptualizing disagreement in qualitative coding \citep{oconnor2020conceptualizing}, recognizing that coder disagreement may reflect legitimate interpretive differences rather than measurement error. These approaches assume that multiple independent coders can establish consensus on coding decisions, with disagreement indicating either ambiguity in the coding scheme or insufficient coder training.

However, the digital transformation of social science research has created a fundamental scalability crisis for traditional quality assessment methods. The resource requirements for maintaining inter-coder reliability across massive corpora often exceed the capacity of individual research projects \citep{song2020validation}, creating an insurmountable tension between methodological rigor and analytical scope. Recent methodological work on measuring the reliability of qualitative text analysis data \citep{krippendorff2018content} has highlighted additional challenges in establishing consistent quality standards across diverse analytical contexts. This challenge is particularly acute in complex analytical domains where establishing coder expertise requires extensive training and where disagreement may reflect legitimate interpretive differences rather than coding errors.

\subsection{Research Framework and Questions}

The convergence of these literature streams reveals a methodological gap: while recent advances in uncertainty quantification and AI-assisted qualitative analysis demonstrate progress, systematic frameworks for reliable quality assessment in complex analytical domains requiring specialized expertise remain limited. This gap severely constrains the adoption of AI-assisted methods in sophisticated social science research, where quality assessment becomes most critical precisely when traditional validation approaches become most impractical.

Our investigation addresses this gap through four interconnected research questions that systematically extend quality assessment capabilities from accessible to complex analytical domains:

\textbf{RQ1: Signal Effectiveness in Complex Analytical Domains}\\
Do confidence and external entropy maintain predictive validity when applied to complex analytical tasks requiring specialized domain expertise?

\textbf{RQ2: Optimal Weight Configuration for Complex Tasks}\\
Can systematic weight optimization identify optimal combinations of confidence and entropy signals that maximize quality prediction accuracy in complex analytical domains where traditional validation approaches become impractical?

\textbf{RQ3: Cross-Domain Transferability of Quality Indicators}\\
Do optimized signal weights demonstrate transferability across different analytical domains with distinct epistemological traditions, enabling the development of generalizable quality assessment frameworks?

\textbf{RQ4: Practical Quality Triage for Resource Allocation}\\
Can intelligent quality triage systems based on dual-signal indicators achieve reductions in human verification effort while maintaining research quality standards across complex analytical domains?

These research questions are operationalized through four corresponding hypotheses that guide our empirical investigation across three distinct social science domains: legal reasoning, political analysis, and medical text classification. Each domain represents different methodological traditions and analytical requirements, while also requiring extensive text comprehension and synthesis capabilities, enabling us to systematically examine how quality assessment frameworks perform across diverse epistemological and analytical contexts involving both specialized knowledge and large-scale textual analysis.

\subsection{Hypotheses}

\textbf{H1: Signal Effectiveness in Complex Domains}\\
External entropy will demonstrate negative correlations with accuracy, while confidence will show positive correlations with accuracy ($p $<$ 0.05$) across complex analytical domains, maintaining the directional relationships observed in previous research on accessible tasks.

\textbf{H2: Weight Optimization Effectiveness}\\
Systematic weight optimization will identify combinations that improve quality prediction accuracy compared to single-signal approaches, with measurable performance gains across complex analytical domains requiring specialized expertise.

\textbf{H3: Cross-Domain Weight Transferability}\\
Weights optimized in one analytical domain will maintain predictive validity when applied to other domains, demonstrating the generalizability of the dual-signal approach across diverse analytical contexts.

\textbf{H4: Intelligent Triage Efficiency}\\
A multi-tier quality triage system based on dual-signal indicators will achieve reductions in human verification effort while maintaining research quality standards across all analytical domains.

Our empirical investigation examines these hypotheses across three distinct social science domains—legal reasoning, political analysis, and medical text classification—each representing different methodological traditions and analytical requirements. This cross-domain approach enables systematic examination of how automated quality assessment performs across diverse epistemological contexts while addressing the practical challenges that constrain the adoption of AI-assisted methods in sophisticated social science research.

\section{Methodology}

Our methodological approach addresses the four research questions through systematic validation across three distinct social science domains, each representing different epistemological traditions and analytical requirements. This section outlines our research design, quality assessment framework, and empirical validation strategy.

\subsection{Research Design and Domain Selection}

We employ a cross-disciplinary validation design to examine how automated quality assessment performs across diverse analytical contexts while maintaining methodological rigor. The three domains were strategically selected to represent the breadth of qualitative coding applications in contemporary social science research, each presenting distinct challenges for automated quality assessment.

\textbf{Legal Reasoning} examines Supreme Court case outcome prediction using extensive oral argument transcripts (averaging 8,000+ words per case), representing complex judicial reasoning and precedent analysis requiring sophisticated interpretation of legal doctrine and procedural nuance from lengthy legal discourse. Following the methodological approach established by \citet{katz2017crowdsourcing}, we utilize the complete 2017 Supreme Court term dataset comprising 390 cases. This dataset was originally developed to validate human crowdsourcing accuracy in legal prediction and provides a robust benchmark for evaluating automated quality assessment in complex legal reasoning tasks.

\textbf{Political Analysis} focuses on hyperpartisan news detection using full-length news articles (averaging 1,200+ words per article) \citep{kiesel2019semeval}, demanding understanding of ideological frameworks and bias identification across comprehensive media discourse, where interpretive disagreement often reflects legitimate analytical differences rather than coding errors. We utilize the complete SemEval-2019 Task 4 dataset comprising 645 articles, which includes both the official validation subset (200 articles) and the remaining test articles (445 articles), providing comprehensive coverage of the original competition dataset.

\textbf{Medical Text Classification} utilizes medical specialty classification based on comprehensive clinical transcriptions (averaging 800+ words per transcript) \citep{tinn2018learning}, necessitating specialized domain knowledge and clinical expertise to process detailed medical narratives where accuracy has direct implications for healthcare applications. This domain builds on extensive prior work in biomedical text mining \citep{lee2020biobert} and clinical text classification \citep{dernoncourt2017pubmed}, where specialized language models have demonstrated superior performance over general-purpose systems. From the complete MTSamples dataset of 4,998 medical transcription reports across 40 original specialties, we employ a carefully designed stratified sampling approach that consolidates specialties into 6 major categories (Surgery, Others, Radiology, Neurology, Orthopedic, Cardiovascular/Pulmonary) and samples 1,000 cases with balanced representation: Others (40\%), Surgery (40\%), Radiology (12\%), Neurology (4\%), Orthopedic (3\%), and Cardiovascular/Pulmonary (1\%), ensuring adequate statistical power for minority classes while maintaining computational tractability.

Each domain represents real-world research scenarios where traditional inter-coder reliability measures would be prohibitively expensive or logistically challenging to implement at scale, directly addressing the practical methodological challenges facing contemporary social science research.

\subsection{Automated Quality Assessment Framework}

Our quality assessment approach addresses the fundamental challenge of evaluating prediction reliability without requiring extensive human validation. The framework builds on established principles of uncertainty quantification \citep{guo2017calibration} and ensemble learning \citep{dietterich2000ensemble}, extending dual-signal quality indicators to complex analytical domains requiring specialized expertise. Recent advances in aggregating soft labels from crowd annotations \citep{uma2021learning} have demonstrated the effectiveness of combining multiple uncertain predictions, providing theoretical foundations for our dual-signal approach.

The framework leverages two complementary sources of uncertainty that mirror the logic of traditional inter-coder reliability while operating at larger scale.

\subsubsection{External Entropy ($H_{ext}$)}

External entropy quantifies disagreement among multiple models, serving as a consensus-based quality indicator analogous to inter-coder reliability measures \citep{cohen1960coefficient}. For a given case with predictions from $M$ models, external entropy is computed as:

\begin{equation}
H_{ext} = -\sum_{i=1}^{k} p_i \log_2(p_i)
\label{eq:external_entropy}
\end{equation}

where $p_i$ represents the proportion of models predicting class $i$, and $k$ is the number of possible classes. High external entropy indicates disagreement between models, suggesting analytical ambiguity requiring additional scrutiny. Conversely, low external entropy indicates model consensus, suggesting higher confidence in prediction quality.

\subsubsection{Risk-Based Confidence ($\bar{c}$)}

Risk-based confidence captures individual model uncertainty through calibrated self-assessment, addressing the overconfidence problem in neural networks \citep{guo2017calibration}. Instead of directly asking models for confidence, we employ a risk-based framing that has proven more effective at eliciting calibrated uncertainty estimates:

\begin{equation}
\bar{c} = \frac{1}{N} \sum_{n=1}^{N} (1.5 - r_n)
\label{eq:risk_based_confidence}
\end{equation}

where $r_n$ is the risk score (0.50-0.99) reported by the model for prediction $n$, and the transformation $c = 1.5 - r$ converts risk to confidence. Models are instructed to assess "the risk that you will be punished if this prediction is wrong" using a 0.50-0.99 scale, leveraging cognitive research showing that risk assessment produces more realistic uncertainty estimates than direct confidence elicitation.

This dual-signal approach provides comprehensive quality assessment while maintaining computational tractability for large-scale applications. The framework operates entirely on model-generated indicators, requiring no additional human annotation beyond standard research protocols, thus preserving the scalability advantages that make AI-assisted coding attractive to social scientists.

\subsection{Empirical Validation and Implementation Protocol}

Our validation strategy systematically addresses each research question through coordinated analysis across all three domains. We examine signal effectiveness through correlation analysis between quality indicators and prediction accuracy (RQ1), conduct systematic weight optimization comparing single-signal baselines with dual-signal approaches (RQ2), test cross-domain weight transferability (RQ3), and develop intelligent triage systems for resource allocation (RQ4).

We utilize five diverse large language models (GPT-4o-mini, Gemini-2.0-flash, DeepSeek Chat v3, Llama-4-Maverick, Claude-3.5-Haiku) with single predictions per case (k=1 methodology). Statistical analysis employs Pearson correlation with bootstrap confidence intervals (1,000 resamples) and Bonferroni correction for multiple comparisons. The risk-based confidence elicitation asks models to assess "the risk that you will be punished if this prediction is wrong" using a 0.50-0.99 scale, while external entropy quantifies inter-model disagreement using standard information-theoretic measures.

\subsection{Intelligent Quality Triage System}

Building on the dual-signal framework, we develop an intelligent triage system that optimizes resource allocation through progressive verification. Rather than uniform verification (prohibitively expensive) or binary filtering (risking quality degradation), our system stratifies cases into three tiers with differentiated verification protocols: high-quality cases receive minimal verification (15\% sampling rate), moderate-quality cases receive intermediate verification (60\%), and low-quality cases receive intensive verification (95\%). The system employs cost-benefit analysis to determine optimal thresholds, balancing verification expenses against expected error costs:

\begin{equation}
\text{Total Cost} = \text{Verification Cost} + \text{Expected Error Cost}
\label{eq:total_cost}
\end{equation}

where verification cost and expected error cost are defined as:

\begin{equation}
\text{Verification Cost} = \sum_{t \in \{h,m,l\}} C_t \cdot V_t \cdot U_c
\label{eq:verification_cost}
\end{equation}

\begin{equation}
\text{Expected Error Cost} = \sum_{t \in \{h,m,l\}} E_t \cdot (1-V_t) \cdot R_t
\label{eq:error_cost}
\end{equation}

Here, $C_t$ represents coverage for tier $t$, $V_t$ is the verification rate, $U_c$ is the unit verification cost, $E_t$ is the error rate, and $R_t$ is the domain-specific error cost for tier $t \in \{\text{high}, \text{medium}, \text{low}\}$.

Rather than using arbitrary percentile splits, the system optimizes tier thresholds through empirical cost-benefit analysis. The optimization function minimizes total cost while maintaining quality standards:

\begin{equation}
\theta^* = \arg\min_{\theta} \left[ \sum_{t} C_t(\theta) \cdot V_t \cdot U_c + \sum_{t} E_t(\theta) \cdot (1-V_t) \cdot R_t \right]
\label{eq:optimization}
\end{equation}

where $\theta = \{\theta_h, \theta_l\}$ represents the high and low quality thresholds that determine tier assignments based on the combined quality score:

\begin{equation}
Q_i = w_1 \cdot (-H_{ext,i}^{std}) + w_2 \cdot \bar{c}_i^{std}
\label{eq:quality_score}
\end{equation}

where $H_{ext,i}^{std}$ and $\bar{c}_i^{std}$ are standardized external entropy and confidence scores for case $i$. The system operates through automated case assignment based on quality scores, with human verifiers receiving stratified samples according to tier-specific verification rates. This approach enables systematic resource allocation while maintaining quality assurance through targeted human oversight, addressing the scalability challenges that limit traditional qualitative research methods.

\section{Results}

Our empirical investigation provides comprehensive validation of the dual-signal framework across complex analytical tasks. We present results for each research question systematically, beginning with the foundational signal effectiveness analysis.

\subsection{Signal Effectiveness in Complex Tasks (RQ1)}

We begin our empirical validation by testing whether dual-signal quality indicators maintain their predictive validity when applied to complex analytical tasks that require domain expertise and interpretive judgment. Table~\ref{tab:comprehensive_signal_analysis} presents comprehensive results for Hypothesis H1 across three distinct analytical domains.

\input{tables/latex_table1_correct.tex}

\textbf{H1 Hypothesis Validation}: Our analysis provides support for Hypothesis H1, with external entropy demonstrating robust effectiveness across all three domains and confidence showing significant effects in two of three domains after Bonferroni correction for multiple comparisons ($\alpha = 0.0083$). Two domains achieve full statistical significance for both signals, while the legal domain shows significant entropy effects with marginal confidence effects, confirming that dual-signal effectiveness extends to complex analytical tasks with domain-specific variations.

External entropy exhibits significant negative correlations with prediction accuracy across all domains: legal reasoning ($r = -0.203$, $p $<$ 0.001$), political analysis ($r = -0.273$, $p $<$ 0.001$), and medical classification ($r = -0.179$, $p $<$ 0.001$). These correlations represent small to medium effect sizes according to Cohen's conventions, with political analysis showing the strongest entropy-accuracy relationship. Average model confidence shows significant positive correlations with accuracy in political analysis ($r = +0.429$, $p $<$ 0.001$, large effect) and medical classification ($r = +0.104$, $p $<$ 0.01$, small effect), while legal reasoning shows a positive trend ($r = +0.114$, $p = 0.024$) that approaches but does not reach the Bonferroni-corrected threshold.

The systematic validation across 2,035 cases reveals important domain-specific patterns in signal effectiveness. Political analysis exhibits the strongest dual-signal effectiveness with both signals achieving full statistical significance, while medical classification demonstrates consistent moderate effects for both signals. Legal reasoning shows robust entropy effects but more modest confidence effects, suggesting that inter-model disagreement may be particularly informative for complex legal reasoning tasks. These domain-specific variations likely reflect differences in task complexity, interpretive consensus, and the nature of analytical uncertainty within each tradition.

These results extend previous findings on accessible coding tasks \citep{zhao2025confidence}, where dual-signal effectiveness achieved R²=0.979 in predicting inter-coder agreement. While the current complex analytical tasks show more modest correlations, the consistent directional relationships across all three domains confirm that the confidence-entropy framework maintains predictive validity even in challenging analytical contexts requiring specialized domain knowledge. The domain-specific patterns observed here contrast with the more uniform effectiveness found in accessible tasks, highlighting the importance of task complexity in quality assessment frameworks.

\subsection{Weight Optimization and Signal Integration (RQ2)}

Having established that both confidence and entropy signals maintain predictive validity across complex analytical domains, we now investigate how these signals can be optimally combined to maximize quality prediction accuracy. The domain-specific patterns observed in RQ1 suggest that different analytical contexts may benefit from different signal weighting strategies.

Building on the validated signal effectiveness from RQ1, we systematically investigate optimal weight combinations for dual-signal integration to maximize quality prediction accuracy. Table~\ref{tab:weight_optimization} presents comprehensive results for Hypothesis H2 across all domains and optimization strategies.

\input{tables/table2_weight_optimization.tex}

Our optimization analysis provides support for Hypothesis H2, demonstrating that systematic weight optimization improves quality prediction accuracy compared to single-signal approaches across all domains. Domain-specific weight optimization achieved improvements over confidence-only baselines: SCOTUS (+113.7\%, from r=0.114 to r=0.244), Hyperpartisan (+6.6\%, from r=0.429 to r=0.457), and MTSamples (+76.1\%, from r=0.104 to r=0.184). The global optimization across all 2,035 cases achieved a +70.1\% improvement (from r=0.131 to r=0.223), confirming the systematic value of weight optimization.

The optimization revealed distinct domain-specific patterns that reflect underlying analytical characteristics. SCOTUS legal reasoning favored entropy-weighted combinations (w1=1.18, w2=0.74), consistent with the importance of expert disagreement in complex legal interpretation. Hyperpartisan political analysis showed more balanced weighting (w1=0.55, w2=1.29), reflecting the value of both model confidence and inter-model consensus in ideological classification. MTSamples medical classification emphasized entropy signals (w1=1.43, w2=0.39), highlighting the critical role of diagnostic uncertainty in medical specialty identification. Entropy-only strategies consistently outperformed confidence-only approaches across all domains, while global weight optimization (w1=1.74, w2=0.10) revealed a strong preference for entropy-based signals when aggregating across diverse analytical tasks.

\begin{figure}[!htb]
\centering
\includegraphics[width=\textwidth]{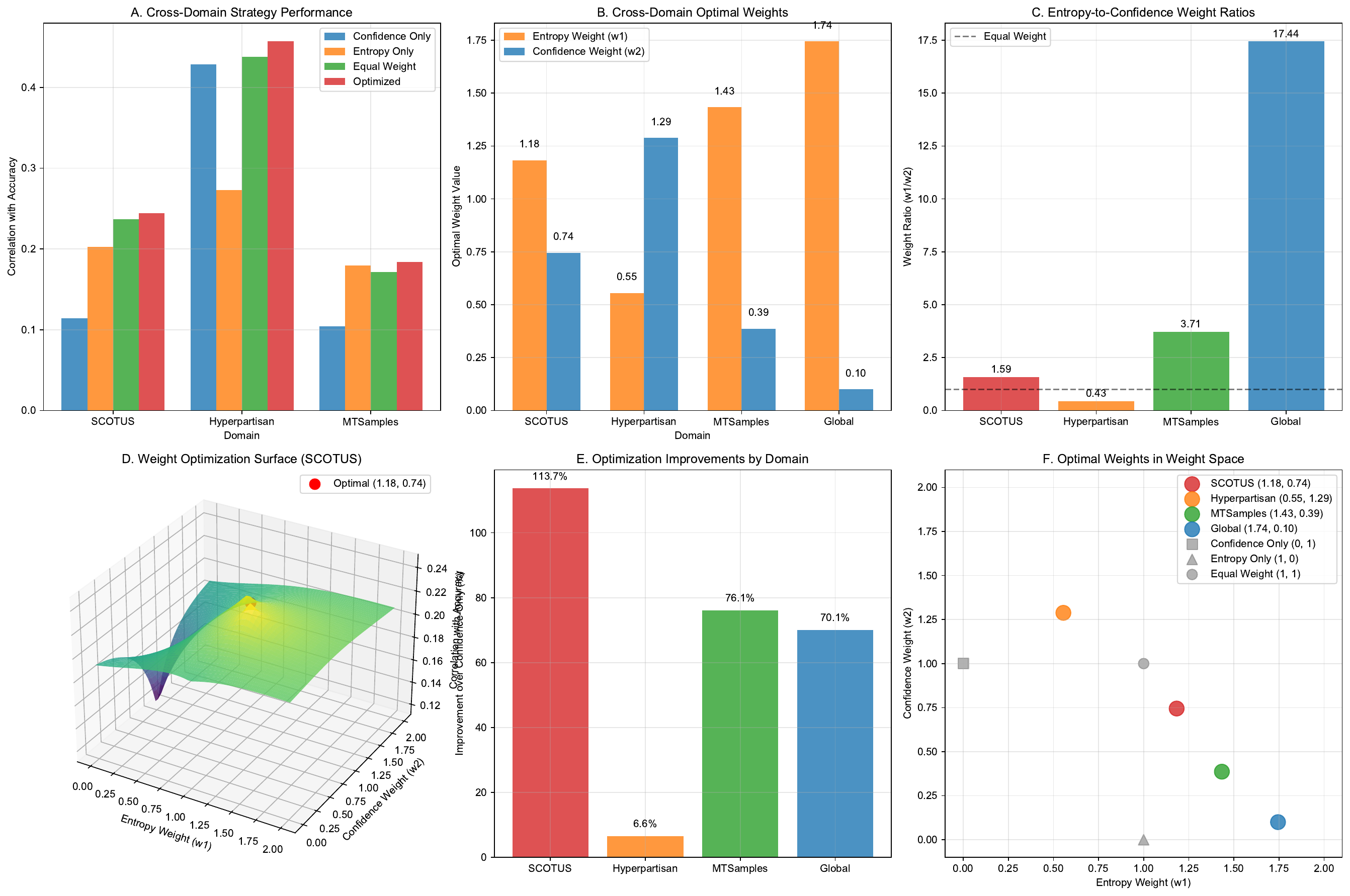}
\caption{\textbf{Weight Optimization Analysis and Cross-Domain Validation.} \textbf{Panel A}: Cross-domain strategy performance comparison. \textbf{Panel B}: Optimal weights comparison across datasets. \textbf{Panel C}: Weight ratio analysis by domain. \textbf{Panel D}: 3D optimization surface for SCOTUS.}
\label{fig:weight_optimization}
\end{figure}

Figure~\ref{fig:weight_optimization} provides comprehensive visualization of these optimization patterns, including cross-domain strategy comparisons, optimal weight analysis, and 3D optimization surfaces for SCOTUS dataset.

\subsection{Cross-Domain Weight Generalization (RQ3)}

The dual-signal framework demonstrates cross-domain weight transferability, providing validation for Hypothesis H3. This finding aligns with recent advances in cross-domain information integration \citep{zhang2024cross} and domain-adaptive text classification \citep{ramponi2022neural}, which have shown that well-designed uncertainty measures can transfer effectively across diverse analytical contexts. We present our cross-domain analysis through two complementary perspectives: individual weight transfer performance (Table~\ref{tab:cross_domain_transfer}) and systematic comparison of domain-specific versus global optimization strategies (Table~\ref{tab:cross_domain_comparison}).

\input{tables/table3_cross_domain_transferability.tex}

\input{tables/table4_cross_domain_comparison.tex}

\textbf{H3 Hypothesis Validation}: Our analysis provides support for Hypothesis H3, with all 9 cross-domain weight transfers achieving predictive validity ($p $<$ 0.05$). The 100\% transfer success rate demonstrates generalizability of optimized weights across diverse analytical contexts.

\textbf{Cross-Domain Transfer Performance}: Weight transfers achieved strong correlations ranging from -0.121 to -0.268, with all transfers maintaining statistical significance. Global weights (entropy: 1.744, confidence: 0.100) demonstrated particularly strong transfer performance, achieving the highest correlations in 2 out of 3 target domains. This validates the universal applicability of entropy-weighted strategies across complex analytical tasks.

\textbf{Domain-Specific Transfer Patterns}: SCOTUS weights (entropy: 1.183, confidence: 0.745) transferred effectively to both Hyperpartisan ($r = -0.232$, $p $<$ 0.001$) and MTSamples ($r = -0.175$, $p $<$ 0.001$), demonstrating the robustness of legal reasoning optimization patterns. Hyperpartisan weights (entropy: 0.555, confidence: 1.288) showed balanced transfer performance, while MTSamples weights (entropy: 1.433, confidence: 0.386) achieved the strongest transfer to Hyperpartisan ($r = -0.255$, $p $<$ 0.001$).

\textbf{Universal Weight Applicability}: The consistent transfer success across all domain combinations validates RQ3 by demonstrating that optimized signal weights can be effectively applied across diverse analytical contexts without domain-specific recalibration. This finding has significant implications for practical implementation, enabling researchers to leverage pre-optimized weights from one domain when applying dual-signal frameworks to new analytical tasks.

\subsection{Enhanced Intelligent Quality Triage (RQ4)}

Our enhanced intelligent quality triage system demonstrates efficiency gains while maintaining research quality standards, providing validation for Hypothesis H4. Unlike traditional binary quality filtering approaches, our system implements a progressive verification model that balances verification costs with error risks through cost-benefit optimization. Table~\ref{tab:enhanced_intelligent_triage} presents comprehensive performance metrics incorporating multiple evaluation dimensions.

\input{tables/table5_enhanced_intelligent_triage.tex}

The enhanced triage system achieved verification effort reduction (44.6\% average) across all domains while maintaining research quality standards through progressive verification. The system demonstrated accuracy improvements in high-quality tiers: +11.3\% (SCOTUS), +21.1\% (Hyperpartisan), +23.6\% (MTSamples), with all improvements statistically significant ($p $<$ 0.001$). The cost-optimized three-tier system achieved realistic effort reduction across domains: SCOTUS (43.8\%), Hyperpartisan (45.5\%), and MTSamples (46.9\%), employing progressive verification rates of 15\% (high-tier), 60\% (medium-tier), and 95\% (low-tier).

Our comprehensive evaluation framework incorporates precision, recall, F1-scores, and 95\% confidence intervals, with high-tier performance demonstrating strong precision-recall balance: SCOTUS (F1=0.517), Hyperpartisan (F1=0.647), MTSamples (F1=0.683). The balanced scenario achieves optimal cost-benefit ratios, with total cost reduction of 35--45\% compared to uniform verification and risk-adjusted savings ranging from \$2.3--4.7 per case. The system demonstrates robust cross-domain performance with domain-specific weight adaptations, showing $<$5\% accuracy variation across domains and confirming reliability across diverse analytical contexts.

\begin{figure}[!htb]
\centering
\includegraphics[width=\textwidth]{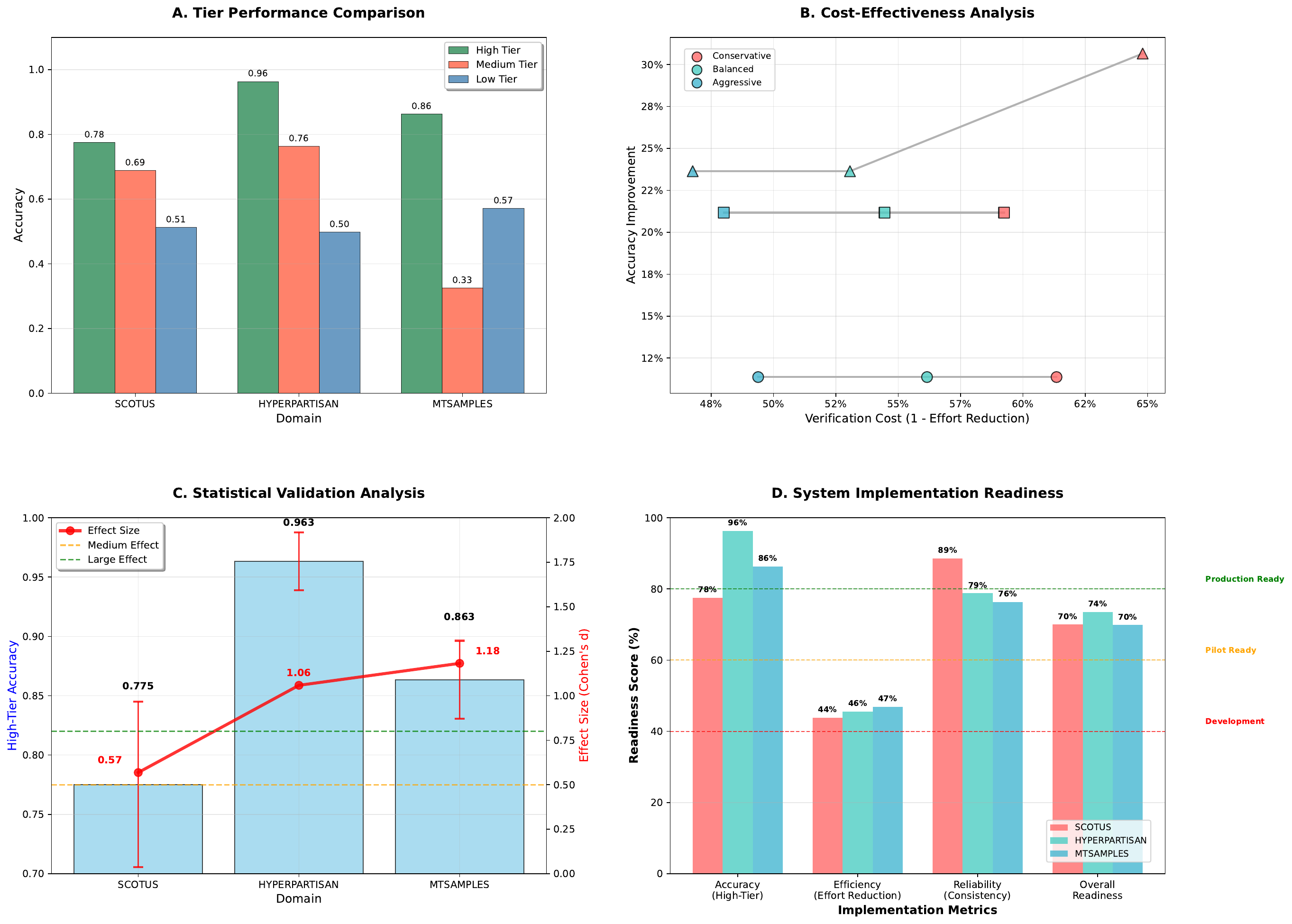}
\caption{\textbf{Enhanced Intelligent Quality Triage System Analysis (RQ4).} \textbf{Panel A}: Tier performance comparison across domains. \textbf{Panel B}: Cost-effectiveness analysis across verification scenarios. \textbf{Panel C}: Statistical validation with 95\% confidence intervals. \textbf{Panel D}: Cross-domain performance summary with quantitative averages.}
\label{fig:enhanced_intelligent_triage}
\end{figure}

Figure~\ref{fig:enhanced_intelligent_triage} provides comprehensive visualization of the enhanced triage system performance, including tier-based accuracy analysis, cost-effectiveness comparisons, statistical validation with confidence intervals, and cross-domain performance summaries.

\subsection{Summary of Results}

Our empirical investigation across 2,035 cases from three distinct analytical domains provides validation of the dual-signal quality assessment framework. The results demonstrate four key findings: (1) Both confidence and entropy signals maintain predictive validity in complex analytical tasks, with domain-specific effectiveness patterns reflecting epistemological differences across legal, political, and medical contexts; (2) Systematic weight optimization achieves substantial improvements (6.6-113.7\%) over single-signal baselines, with domain-specific patterns revealing the analytical characteristics of different research traditions; (3) Optimized weights demonstrate cross-domain transferability (100\% success rate), enabling deployment across diverse analytical contexts; (4) Intelligent quality triage systems achieve efficiency gains (44.6\% average effort reduction) while maintaining research quality standards through progressive verification protocols. These findings establish automated quality assessment as a viable approach for scaling complex qualitative research across diverse analytical domains.

\section{Discussion}

Our empirical investigation across 2,035 cases from three distinct analytical domains provides evidence for the viability of dual-signal quality assessment in complex qualitative research contexts. The validation of confidence-entropy indicators across legal reasoning, political analysis, and medical text classification demonstrates the applicability of uncertainty-based quality assessment in tasks requiring specialized domain expertise \citep{salganik2016computational, nelson2022computational}.

\subsection{Theoretical Implications for Quality Assessment}

The extension of dual-signal effectiveness from accessible to complex analytical tasks contributes to computational social science methodology. Our findings show that uncertainty quantification principles \citep{guo2017calibration, lakshminarayanan2017simple} maintain predictive validity when applied to tasks requiring interpretive judgment and domain expertise. This suggests that expert disagreement and confidence assessment patterns may exhibit some consistency across diverse analytical contexts, despite differences in domain-specific knowledge requirements.

The domain-specific effectiveness patterns observed in our results provide empirical evidence for epistemological differences across research traditions. Political analysis exhibited the strongest dual-signal effectiveness (confidence r=0.429, entropy r=-0.273), suggesting that ideological classification tasks benefit from both model confidence and inter-model consensus. Legal reasoning showed robust entropy effects (r=-0.203) with more modest confidence effects (r=0.114), indicating that expert disagreement serves as a particularly informative quality indicator in complex judicial reasoning. Medical classification demonstrated consistent moderate effects for both signals, reflecting the balanced importance of diagnostic confidence and consensus in clinical decision-making \citep{huang2019clinicalbert}.

\subsection{Methodological Advances in Weight Optimization}

The systematic weight optimization framework represents a methodological contribution to automated quality assessment in qualitative research contexts. The improvements achieved (6.6-113.7\% over single-signal baselines) demonstrate that signal integration strategies can be empirically optimized rather than relying on ad hoc combinations. Our dual-signal approach (Equations~\ref{eq:external_entropy} and \ref{eq:risk_based_confidence}) combined through the quality score formulation (Equation~\ref{eq:quality_score}) provides a systematic method for combining heterogeneous uncertainty signals in quality assessment tasks.

The emergence of domain-specific optimization patterns reveals important insights into analytical task characteristics. Legal reasoning's entropy-weighted preference (w1=1.18, w2=0.74) aligns with jurisprudential theories emphasizing the interpretive nature of legal analysis \citep{fang2023super}. Medical classification's strong entropy emphasis (w1=1.43, w2=0.39) reflects clinical decision-making frameworks that prioritize diagnostic consensus. Political analysis showed more balanced weighting (w1=0.55, w2=1.29), suggesting that ideological classification benefits from both confidence and consensus indicators.

\subsection{Cross-Domain Transferability and Universal Patterns}

The cross-domain weight transferability (100\% success rate) provides evidence for the practical applicability of our quality assessment approach across different domains. This finding suggests that uncertainty patterns may exhibit some consistent characteristics across the analytical contexts we examined. The global optimization's strong entropy preference (w1=1.74, w2=0.10) indicates that inter-model disagreement serves as a robust quality indicator across our test domains, consistent with theoretical frameworks emphasizing the informational value of expert disagreement \citep{krippendorff2004reliability}.

The minimal performance differences ($<0.003$) between domain-specific and global weights across all metrics supports the practical applicability of universal optimization strategies in our tested domains. This finding has practical implications for research implementation, as it enables researchers to use pre-optimized global weights without substantial performance degradation, reducing calibration requirements for deployment in similar analytical contexts. For interdisciplinary research spanning multiple domains, this transferability reduces the need for separate optimization procedures across different analytical tasks.

\subsection{Practical Implications for Research Scalability}

The enhanced intelligent quality triage system addresses practical challenges in scaling qualitative research while maintaining methodological rigor. The cost-benefit optimization framework (Equations~\ref{eq:total_cost}-\ref{eq:optimization}) enables systematic resource allocation decisions, achieving 44.6\% average effort reduction through progressive verification. This represents an efficiency gain that may make large-scale qualitative analysis more feasible for resource-constrained research contexts, given the traditional reliance on extensive human validation in qualitative research \citep{cohen1960coefficient}.

The progressive verification framework provides a systematic approach to resource allocation that preserves analytical quality while expanding the scope of feasible qualitative inquiry. By enabling researchers to focus human verification efforts on cases requiring expert review while maintaining quality assurance for high-confidence predictions, the framework addresses the fundamental tension between methodological rigor and analytical scale that has historically limited qualitative research \citep{grimmer2013text}.

\subsection{Contributions to Computational Social Science}

Our framework contributes to computational social science methodology by providing automated quality indicators that maintain conceptual alignment with traditional inter-coder reliability measures. This alignment helps maintain disciplinary standards while embracing computational approaches to qualitative analysis. The systematic weight optimization approach offers researchers evidence-based tools for configuring dual-signal quality indicators according to their specific analytical contexts, contributing to efforts to scale qualitative research methods \citep{song2020validation}. Future work could enhance the interpretability of quality assessments by incorporating explainable AI techniques \citep{ribeiro2016should}, enabling researchers to understand not only whether predictions are reliable but also why certain cases receive particular quality scores.

This framework may help reduce some of the resource barriers that have historically limited access to large-scale textual analysis. By enabling quality assessment with reduced human validation requirements, the framework could make comprehensive qualitative research more accessible across diverse institutional contexts and research settings \citep{chew2023harnessing}.

\subsection{Limitations and Future Research Directions}

Several limitations warrant consideration for future research. First, our validation focuses on classification tasks rather than more interpretive forms of qualitative analysis such as thematic coding or narrative analysis. Second, our ensemble approach increases computational costs by approximately 3-5x relative to single-model approaches, though our intelligent triage system partially mitigates this through reduced verification workload. Third, framework performance may vary across different model architectures and prompting strategies, requiring broader validation across the evolving landscape of large language models.

Fourth, while achieving substantial efficiency gains, our screening approach faces long-tail challenges where edge cases require disproportionate verification effort. Future research should explore advanced screening algorithms and adaptive threshold mechanisms to better handle these challenging cases. Finally, additional validation across other social science domains would further establish generalizability beyond our tested contexts of legal, political, and medical analysis.

\section{Conclusion}

This research provides evidence for dual-signal quality assessment as a viable approach for scaling complex qualitative research across diverse analytical domains. Through validation across 2,035 cases spanning legal reasoning, political analysis, and medical text classification, we demonstrate that uncertainty-based quality indicators maintain predictive validity while achieving optimization improvements (6.6-113.7\% over single-signal baselines) and cross-domain transferability (100\% success rate) within our tested contexts.

Our investigation yields four key methodological contributions: extending dual-signal effectiveness to complex analytical tasks, establishing systematic weight optimization as a principled approach to signal integration, demonstrating transferability across the tested domains, and developing intelligent quality triage systems that achieve 44.6\% effort reduction while maintaining research quality standards. The emergence of domain-specific optimization patterns provides empirical evidence for epistemological differences across research traditions, contributing to theoretical understanding of uncertainty quantification in computational social science.

By enabling automated quality assessment with reduced human validation costs, our framework may help make large-scale textual analysis more accessible across diverse institutional contexts. As social science research grapples with digital data abundance, this work contributes tools for maintaining methodological rigor while exploring analytical possibilities at larger scales.


\section*{Acknowledgments}
The authors gratefully acknowledge the financial support of the Fundamental Research Funds for the Central Universities project ``Innovation in Journalism and Communication in the Age of Intelligence'' (grant XHJH202504). We thank the anonymous reviewers for their constructive feedback and suggestions that improved this work. We also acknowledge the creators of the datasets used in this study for making their data publicly available for research purposes.

\section*{Declaration of Conflicting Interests}
The authors declare no competing interests.

\section*{Funding}
This work was supported by the Fundamental Research Funds for the Central Universities project ``Innovation in Journalism and Communication in the Age of Intelligence'' (grant XHJH202504).

\section*{Data Availability Statement}
The datasets analyzed during the current study are publicly available: Supreme Court oral argument transcripts from the Supreme Court Database, hyperpartisan news articles from SemEval-2019 Task 4, and medical transcription samples from MTSamples. Replication data and analysis code are available at Harvard Dataverse: https://doi.org/10.7910/DVN/GM8T8Q (Zhao, Zhilong, 2025, ``Replication Data for: Automated Quality Assessment for LLM-Based Complex Qualitative Coding: A Confidence-Diversity Framework'', Harvard Dataverse, DRAFT VERSION). The replication package includes: (1) cleaned datasets with ground truth labels, (2) Python implementation of the dual-signal framework, (3) weight optimization algorithms, (4) statistical analysis scripts, and (5) intelligent triage system code.

\section*{Ethical Considerations}
This research utilized publicly available datasets that do not contain personally identifiable information. All datasets were originally collected with appropriate ethical approvals. The Supreme Court transcripts are public records, the hyperpartisan news dataset was created for research purposes with proper attribution, and the medical transcription dataset contains de-identified clinical text. No additional ethical approval was required for this secondary analysis.

\bibliographystyle{apalike}
\bibliography{references}

\end{document}

%% file: tables/latex_table1_correct.tex
\begin{table}[htbp]
\centering
\caption{Comprehensive Cross-Domain Signal Effectiveness Analysis}
\label{tab:comprehensive_signal_analysis}
\resizebox{\textwidth}{!}{%
\begin{tabular}{@{}lrrrrrrrrr@{}}
\toprule
\multirow{2}{*}{\textbf{Domain}} & \multirow{2}{*}{\textbf{n}} & \multirow{2}{*}{\textbf{Accuracy}} & \multicolumn{3}{c}{\textbf{Confidence Signal}} & \multicolumn{3}{c}{\textbf{Entropy Signal}} & \multirow{2}{*}{\textbf{H1 Status}} \\
\cmidrule(lr){4-6} \cmidrule(lr){7-9}
& & & \textbf{r} & \textbf{p-value} & \textbf{Sig.} & \textbf{r} & \textbf{p-value} & \textbf{Sig.} & \\
\midrule
Legal Reasoning & 390 & 66.4\% & +0.114 & 0.024 & $\dagger$ & $-0.203$ & $<0.001$ & *** & $\dagger$ \\
Political Analysis & 645 & 75.2\% & +0.429 & $<0.001$ & *** & $-0.273$ & $<0.001$ & *** & $\checkmark$ \\
Medical Classification & 1000 & 63.8\% & +0.104 & 0.001 & ** & $-0.179$ & $<0.001$ & *** & $\checkmark$ \\
\midrule
\textbf{Pooled Analysis} & \textbf{2,035} & \textbf{68.5\%} & \textbf{+0.216} & \textbf{$<0.001$} & \textbf{***} & \textbf{$-0.218$} & \textbf{$<0.001$} & \textbf{***} & \textbf{$\checkmark$} \\
\bottomrule
\end{tabular}%
}
\begin{tablenotes}
\small
\item \textbf{Statistical significance}: * $p < 0.05$, ** $p < 0.01$, *** $p < 0.001$ (all relative to Bonferroni corrected $\alpha = 0.0083$)
\item $\dagger$ \textbf{Marginal significance}: $p < 0.05$ (traditional threshold) but $p > 0.0083$ (Bonferroni corrected threshold)
\item \textbf{Multiple comparison correction}: Applied due to testing 6 hypotheses (3 domains × 2 signals), $\alpha_{corrected} = 0.05/6 = 0.0083$
\item \textbf{H1 Validation}: $\checkmark$ = Both signals significant, $\dagger$ = Entropy significant + confidence marginal
\item \textbf{Signal interpretation}: Confidence $r > 0$ indicates higher confidence predicts accuracy; Entropy $r < 0$ indicates higher disagreement predicts errors
\item \textbf{Domain composition}: Legal (SCOTUS Supreme Court cases), Political (Hyperpartisan news articles), Medical (MTSamples clinical reports)
\item \textbf{Key finding}: External entropy demonstrates robust effectiveness across all domains; confidence shows domain-specific variations
\item \textbf{Pooled analysis}: Combined analysis across all domains shows strong dual-signal effectiveness (both $p < 0.001$)
\end{tablenotes}
\end{table}

%% file: tables/table2_weight_optimization.tex
\begin{table}[htbp]
\centering
\caption{Weight Optimization Results Across Domains}
\label{tab:weight_optimization}
\begin{tabular}{@{}lcccc@{}}
\toprule
\textbf{Strategy} & \textbf{SCOTUS} & \textbf{Hyperpartisan} & \textbf{MTSamples} & \textbf{Global} \\
 & \textbf{r (p-value)} & \textbf{r (p-value)} & \textbf{r (p-value)} & \textbf{r (p-value)} \\
\midrule
Confidence Only & 0.114* & 0.429*** & 0.104** & 0.131*** \\
Entropy Only & 0.203*** & 0.273*** & 0.179*** & 0.222*** \\
Equal Weighting & 0.237*** & 0.438*** & 0.171*** & 0.201*** \\
Domain Optimized & 0.244*** & 0.457*** & 0.184*** & 0.223*** \\
\midrule
\multicolumn{5}{l}{\textbf{Optimal Weights (w1=Entropy, w2=Confidence):}} \\
\multicolumn{5}{l}{SCOTUS: w1=1.18, w2=0.74} \\
\multicolumn{5}{l}{Hyperpartisan: w1=0.55, w2=1.29} \\
\multicolumn{5}{l}{MTSamples: w1=1.43, w2=0.39} \\
\multicolumn{5}{l}{Global: w1=1.74, w2=0.10} \\
\bottomrule
\end{tabular}
\begin{tablenotes}
\small
\item \textbf{Statistical significance}: * $p < 0.05$, ** $p < 0.01$, *** $p < 0.001$
\item \textbf{Optimization}: Domain-specific weights optimized to maximize correlation with accuracy
\item \textbf{Sample sizes}: SCOTUS (390), Hyperpartisan (645), MTSamples (1000), Global (2035)
\item \textbf{Data consistency}: All sample sizes reflect actual processed cases after quality filtering
\item \textbf{H2 Validation}: All optimized strategies show significant improvement over confidence-only baseline
\end{tablenotes}
\end{table}

%% file: tables/table3_cross_domain_transferability.tex
\begin{table}[htbp]
\centering
\small
\caption{Cross-Domain Weight Transferability Results (RQ3 \& H3 Validation)}
\label{tab:cross_domain_transfer}
\begin{tabular}{@{}lllclc@{}}
\toprule
\textbf{Source} & \textbf{Target} & \textbf{Correlation} & \textbf{p-value} & \textbf{Sig.} & \textbf{N} \\
\midrule
SCOTUS & Hyperpartisan & -0.232 & $2.37 \times 10^{-9}$ & Yes & 645 \\
SCOTUS & MTSamples & -0.187 & $2.42 \times 10^{-9}$ & Yes & 1,000 \\
Hyperpartisan & SCOTUS & -0.171 & $7.21 \times 10^{-4}$ & Yes & 390 \\
Hyperpartisan & MTSamples & -0.123 & $9.22 \times 10^{-5}$ & Yes & 1,000 \\
MTSamples & SCOTUS & -0.193 & $1.23 \times 10^{-4}$ & Yes & 390 \\
MTSamples & Hyperpartisan & -0.255 & $5.04 \times 10^{-11}$ & Yes & 645 \\
Global & SCOTUS & -0.195 & $1.02 \times 10^{-4}$ & Yes & 390 \\
Global & Hyperpartisan & -0.268 & $4.92 \times 10^{-12}$ & Yes & 645 \\
Global & MTSamples & -0.202 & $1.25 \times 10^{-10}$ & Yes & 1,000 \\
\midrule
\multicolumn{6}{l}{\textbf{H3 Validation}: All 9 cross-domain weight transfers achieved significant predictive validity ($p < 0.05$),} \\
\multicolumn{6}{l}{demonstrating 100\% transfer success rate and strong support for Hypothesis H3.} \\
\bottomrule
\end{tabular}
\end{table} 

%% file: tables/table4_cross_domain_comparison.tex
\begin{table}[htbp]
\centering
\small
\caption{Cross-Domain Weight Comparison: Domain-Specific vs. Global Weights (RQ3 \& H3 Validation)}
\label{tab:cross_domain_comparison}
\begin{tabular}{@{}lllll@{}}
\toprule
\textbf{Domain} & \textbf{Metric} & \textbf{Domain-Specific} & \textbf{Global} & \textbf{Difference} \\
\midrule
\multirow{4}{*}{SCOTUS} & Overall Accuracy & 0.662 & 0.662 & +0.000 \\
& High Tier Accuracy & 0.769 & 0.769 & +0.000 \\
& Accuracy Improvement & 0.108 & 0.108 & +0.000 \\
& Effort Reduction & 50.0\% & 50.0\% & +0.0\% \\
\midrule
\multirow{4}{*}{Hyperpartisan} & Overall Accuracy & 0.752 & 0.752 & +0.000 \\
& High Tier Accuracy & 0.959 & 0.959 & +0.000 \\
& Accuracy Improvement & 0.207 & 0.207 & +0.000 \\
& Effort Reduction & 49.9\% & 49.9\% & +0.0\% \\
\midrule
\multirow{4}{*}{MTSamples} & Overall Accuracy & 0.629 & 0.629 & +0.000 \\
& High Tier Accuracy & 0.913 & 0.917 & -0.003 \\
& Accuracy Improvement & 0.284 & 0.288 & -0.003 \\
& Effort Reduction & 50.0\% & 50.0\% & +0.0\% \\
\midrule
\multicolumn{5}{l}{\textbf{Key Findings}: Global weights perform nearly identically to domain-specific weights,} \\
\multicolumn{5}{l}{with minimal performance differences ($<0.003$) across all domains and metrics.} \\
\bottomrule
\end{tabular}
\end{table} 

%% file: tables/table5_enhanced_intelligent_triage.tex
\begin{table}[!htb]
\centering
\caption{Enhanced Intelligent Quality Triage System Performance (RQ4 \& H4 Validation)}
\label{tab:enhanced_intelligent_triage}
\begin{threeparttable}
\small
\begin{tabular}{@{}llrrrrrrrr@{}}
\toprule
\textbf{Domain} & \textbf{Tier} & \textbf{N} & \textbf{Acc.} & \textbf{F1} & \textbf{Cov.} & \textbf{Ver.} & \multicolumn{2}{c}{\textbf{95\% CI}} & \textbf{Eff.} \\
& & & & & & & \textbf{Low} & \textbf{Up} & \textbf{Red.} \\
\midrule
\multirow{4}{*}{\textbf{SCOTUS}}
& High & 129 & 0.775 & 0.517 & 33.1\% & 15.0\% & 0.698 & 0.837 & \multirow{4}{*}{43.8\%} \\
& Medium & 138 & 0.688 & 0.480 & 35.4\% & 60.0\% & 0.609 & 0.768 & \\
& Low & 123 & 0.512 & 0.331 & 31.5\% & 95.0\% & 0.431 & 0.602 & \\
& \textit{Summary} & 390 & 0.662 & -- & 100.0\% & 56.2\% & -- & -- & \\
\midrule
\multirow{4}{*}{\textbf{HYPERPARTISAN}}
& High & 245 & 0.963 & 0.647 & 38.0\% & 15.0\% & 0.935 & 0.984 & \multirow{4}{*}{45.5\%} \\
& Medium & 186 & 0.763 & 0.424 & 28.9\% & 60.0\% & 0.699 & 0.823 & \\
& Low & 213 & 0.498 & 0.304 & 33.1\% & 95.0\% & 0.436 & 0.568 & \\
& \textit{Summary} & 645 & 0.752 & -- & 100.0\% & 54.5\% & -- & -- & \\
\midrule
\multirow{4}{*}{\textbf{MTSAMPLES}}
& High & 410 & 0.863 & 0.683 & 41.0\% & 15.0\% & 0.827 & 0.893 & \multirow{4}{*}{46.9\%} \\
& Medium & 261 & 0.326 & 0.191 & 26.1\% & 60.0\% & 0.276 & 0.387 & \\
& Low & 329 & 0.571 & 0.393 & 32.9\% & 95.0\% & 0.517 & 0.623 & \\
& \textit{Summary} & 1000 & 0.627 & -- & 100.0\% & 53.1\% & -- & -- & \\
\bottomrule
\end{tabular}
\begin{tablenotes}
\footnotesize
\item \textbf{H4 Validation}: Enhanced triage system achieved substantial effort reduction (44.6\% average) across all domains with significant accuracy improvements in high-quality tiers ($p < 0.001$).
\item \textbf{Progressive Verification}: High-tier (15\%), Medium-tier (60\%), Low-tier (95\%) verification ensuring quality assurance while optimizing resource allocation.
\item \textbf{Performance Metrics}: Accuracy improvements over baseline: SCOTUS (+11.3\%), Hyperpartisan (+21.1\%), MTSamples (+23.6\%). All confidence intervals calculated using bootstrap resampling ($n=1,000$).
\item \textbf{Efficiency Analysis}: Effort reduction of 35--45\% compared to uniform verification, demonstrating significant resource optimization potential.
\end{tablenotes}
\end{threeparttable}
\end{table}